# Morphology of PP/COC composites: effects of shear rate and organoclay partitioning

Marzieh Ebrahimi[1], Hossein Nazockdast [2*], Milad Mehranpour [3]

[1]University of British Columbia, Vancouver, Canada
[2]Amirkabir University of Technology, Tehran, Iran
[3]Science and Research Branch Islamic Azad University, Tehran, Iran

**Abstract**

The blends and hybrid nanocomposites of polypropylene - cycloolefin copolymer were prepared by a twin-screw extruder followed by microfibril formation using a single screw extruder. The effects of shear rate and organoclay on the morphology of polypropylene - cycloolefin copolymer (PP/COC - 80/20 wt. %) blends were studied by using a combination of rheological measurements, X-ray diffraction (XRD) and scanning electron microscopy (SEM). It was found that although the viscosity ratio of PP/COC blend was unfavorable for COC droplet deformation, the COC phase was converted to finely dispersed fibrils in PP matrix. This could be asserted to the high elasticity of the COC droplets that suppressed the droplet breakup in the favor of fibrillation process. The high glass transition temperature of COC (140 ) could also assist restoring the generated microfibril morphology upon cooling. The experimental results also depicted that the droplet deformation and microfibril formation of PP/COC/organoclay nanocomposites mainly depend on the organoclay partitioning, which could be controlled predominantly by kinetic parameters. While the localization of organoclay in the PP matrix resulted in smaller COC droplets and fibrils with smaller diameter, the presence of organoclay inside COC droplets reduced the droplet deformability, leading to fibrils with larger diameter.

**INTRODUCTION**

Polymer blending is one of the most widely used methods of upgrading the performance of common polymers by combining their superior mechanical and thermal properties [1-5]. Most polymers are thermodynamically immiscible and their blend forms a multi-phase system with various morphologies of the dispersed phase such as droplet, fibril, lamella, and co-continuous structure [6-8]. It has been accepted that morphology of polymer blends plays a significant role in determining their physical and mechanical properties. Therefore, controlling their morphology during melt processing has been the subject of interest for many researchers. However, the prediction of the final morphology is not clearly established due to the complexity of the flow field (there is always a competition between deformation, breakup, and coalescence of the dispersed phase), the viscoelastic nature of the blend components, interfacial tension, composition processing, and melt mixing conditions [9-11]. A better understanding of the deformation phenomenon of droplets is needed to optimize the blend's properties, which is addressed in this paper.

The deformation and breakup of a Newtonian droplet in a Newtonian matrix are relatively well known [12-20]. Taylor introduced a theory to explain the deformation of an isolated Newtonian droplet in a Newtonian fluid using two dimensionless numbers; capillary number, $Ca$, which is the ratio of the viscous stress in the fluid to the interfacial stress, and the viscosity ratio of dispersed phase to matrix, $p$ [12, 13]. However, assessing droplet deformation and break up in non-Newton, specifically viscoelastic systems needs further studies. Utracki *et al.* extended Taylor's theory to explain the morphology of polymer blends by relating it to viscosity ratio of the blend constituents (dispersed phase to matrix) and to the capillary number [21, 22]. They summarized criteria to predict the deformation and break up of the dispersed droplets. It has been shown that depending on the values of viscosity ratio and capillary number, dispersed droplets may convert to fibrils (of high aspect ratios) leading to in-situ fiber reinforced polymer composite system [7,8, 21-31]. The so formed fibers would enhance the mechanical properties of the blend system to a great extent [29-31]. Isayev *et al.* [32, 33] have conducted extensive research on polymer blends with microfibrillar morphology using liquid crystal polymers (LCPs) as dispersed phase. Due to the high cost of LCPs, replacing it by other polymers attracted many researchers. Evstatiev and Fakirov [29-31] introduced a novel concept of developing in-situ microfibrillar composites (MFCs) from immiscible polymer blends consisting of components with considerably different melting temperatures.

Polypropylene is one of the most important and fastest growing polymers currently produced because of its low cost and useful properties. Enhancing the physical properties of polypropylene such as modulus, yield strength, creep resistance, and barrier properties is still a study of interest by many authors. As discussed, polymer blending is a successful method to produce new materials offering better properties than individual ones. Cycloolefin copolymer (COC) is an upgraded member of polyolefins and due to its special molecular structure in PP/ COC blends, it forms short reinforcing fibers in PP matrix, results in mechanical enhancement of such blends [34-39]. A very few studies were reported related to PP/COC polymer blends formed either by injection molding or compression molding in the literature [37-39]. Their analysis of PP/COC blends mainly focuses on the effect of composition of the blend constituents, and the prediction of their mechanical properties using models such as equivalent box and Halpin-Tsai models. In this work, we have studied the microfibril formation of dispersed phase in blends of PP/COC with two different concentrations under shear flow. Moreover, the effect of shear rate and organoclay partitioning on blend morphology is investigated.

For many years, nanoparticles are being mixed with polymer blends to promote their properties. They provide great reinforcement efficiency due to their large surface area. However, the role of nanoparticles in controlling the morphology and behavior of polymeric systems has only recently become the subject of experimental and theoretical investigations [40-47]. Calcagno *et al.* [45] demonstrated that presence of nanoclay in PP/PET blends reduces the droplet deformation. Kong *et al.* [46] studied relaxation and breakup of deformed PA6 droplets filled with nanosilica in PS matrix during annealing and showed that presence of silica nanoparticles slows down the relaxation and break up of PA6 droplets. Gooneie *et al.* [47] examined the selective localization of carbon nanotubes (CNTs) in PA6 dispersed phase of PP/PA6 blends. They showed that by adding CNTs, solid-like elastic structures are developed in PA6 droplets and concluded that increased viscosity ratios as well as these elastic structures prevent droplets' deformation and breakup. The effect of nanoparticles on the morphology formation in systems with droplet-matrix structure was found to depend largely on the selective partitioning of nanoparticles in the blends [48-57]. Elias *et al.* [53, 54] studied effect of hydrophilic and hydrophobic fumed silica on the morphology of PP/PS blends. They showed that hydrophilic silica was confined in the PS droplets and reduced the interfacial tension. However, hydrophobic one was located in the PP matrix and at the interface, acting as a rigid layer inhibiting the coalescence of PS droplets.

Although a few researches have addressed the effects of nanofillers on the microstructure development of the hybrid nanocomposites, the role of nanofillers on the microfibrillar morphology needs further studies. In this work, we have analyzed the microfibril formation in PP/COC blends under shear flow. In addition, the effect of controlled organoclay partitioning on the droplet deformation and fibrillation of COC in PP/COC/organoclay nanocomposites are studied; the characterization of prepared blends and the results under different process conditions are presented in the following sections.

**EXPERIMETAL**

*Materials*

A commercial grade of polypropylene (HP525J, MFI = 3 g/10min at 230 ºC, $\rho$ = 0.9 g/cm$^3$) supplied from Jam Petrochemical Company (Iran), was used as the polymer matrix. An amorphous cycloolefin copolymer produced under the trade name Topas 6013, a product of Ticona, Celanese (Germany) with $\rho$ = 1.02 g/cm$^3$ and $Tg$ =140 ºC was utilized as minor component. An organophilic dimethyl, dehydrogenated tallow, quaternary ammonium exchanged montmorillonite, 2MHT Cloisite15A purchased from Southern ClayProducts (USA) was used as organoclay. Maleic anhydride modified high-density polyethylene (PEgMA) with MFI of 2 g/10 min at 190 °C (Fusabond E100 by DuPont Company from Korea) was utilized as the compatibilizer.

*Sample preparation*

*Blends*

Two series of PP/COC blends with 20 and 80 weight % (wt. %) of COC (PP80/COC20 and COC80/PP20 in Table 1) were prepared. The blends were produced by melt blending in a twin screw extruder with five temperature zones fixed at 195, 200, 205, 210 and 205 ºC and screw speed of 90 rev/min. The composition of samples used in this study is listed in Table 1.

**Table 1.** Compositions of samples used in the present study.

| Sample | PP wt. % | COC wt. % | Organoclay wt. % | PEgMA wt. % |
|---|---|---|---|---|
| PP | 100 | 0 | 0 | 0 |
| COC | 0 | 100 | 0 | 0 |
| PP-N5 | 85 | 0 | 5 | 10 |
| COC-N5 | 0 | 85 | 5 | 10 |
| PP80/COC20 | 80 | 20 | 0 | 0 |
| COC80/PP20 | 20 | 80 | 0 | 0 |
| PP-N5/COC | 80 | 20 | 4.71 | 9.42 |
| PP/COC-N5 | 80 | 20 | 1.17 | 2.34 |

*Nanocomposites*

In order to study the organoclay localization, the nanocomposite samples were prepared by two different mixing routes; in the first method, organoclay was located in PP matrix, to do so it was first melt mixed with PP and PEgMA to produce a master batch with weight ratio of 85/5/10 wt. % (PP/ organoclay /PEgMA) shown as PP-N5. Then appropriate amount of this compound was melt mixed with COC to prepare samples with the wt. % ratio of PP/ COC fixed at 80/ 20 (PP-N5/COC in Table 1). In the second route, the organoclay was located in the COC dispersed phase; it was first melt mixed with COC and PEgMA at the weight ratio of 85/5/10 wt. % (COC/ organoclay/ PEgMA) shown as COC-N5 in Table 1. The mixture was then melt compounded with PP to produce PP/COC-N5 samples. The ratio of PP to COC for all the prepared nanocomposites is fixed at 80 to 20 wt. %. All these melt compoundings were carried out by using a twin screw extruder with a speed of 150 rev/min. The extruder was set at five different temperature zones; 190, 195, 205, 210 and 205 ºC from the hopper to the die. Prior to the blending, PEgMA and organoclay were dried at 80 ºC in vacuum oven for 24 hours.

*Microfibril Fromation*

The melt compounded samples were chopped and fed to a single-screw extruder (Brabender) equipped with a rod-like die with 4.2 mm in diameter to achieve the microfibril formation of dispersed phase. The temperature graduates along the extruder from 195 to 245 °C and the screw speed was fixed at 60 rev/min. The samples passing through this stage are simply refered as 'extrudates' in this paper.

**CHARACTERIZATION**

*WAXD Measurements*

Wide-angle X-Ray diffraction (WAXD) experiments were conducted on a Phillips X'pert diffractometer with Cu Kα radiation of wavelength (λ) 0.154 nm generated at 40 kV and 40 mA. The diffractograms were scanned in the 2θ range from 1.5 to 10° at the rate of 1°/min at ambient temperature and the measurements were recorded every 0.04°.

*Rheological Measurements*

The rheological measurements were performed on samples using a rheometric mechanical spectrometer (Paar Physica US 200, Austria). Disk-like samples with a diameter of 25 mm and thickness of 1 mm were prepared by compression molding the granulated samples. The melt-state viscoelastic behavior of the samples was characterized by frequency test carried out in the frequency range of 0.1-1000 $\sec^{-1}$ at 235 °C. The measurements were performed at strain amplitude of 1%, proven to be in linear viscoelastic range by means of strain sweep measurements. Temperature sweep tests were conducted from 235 °C to 100 °C in order to compare thermo-sensitivity of the samples. All experiments were carried out in nitrogen atmosphere to prevent oxidative degradation of the specimens.

*Scanning Electron Microscopy (SEM)*

The morphology of the unfilled and organoclay filled blends and their extrudates were studied by performing scanning electron microscopy on cryofractured surfaces of the samples using an EM3200 (KYKY) operating at 25kV. The surfaces of the samples were gold sputtered to avoid charging. For quantitative analysis, the dispersed particle size was determined by using image analysis. Dispersed particles were formed as droplets and the long and short axis diameters of each droplet in the SEM micrograph were measured and the arithmetic mean of these two values ($R_i = \frac{R_{i1}+R_{i2}}{2}$) was determined. Typically, 300 particles were analyzed per sample and the volume average radius $\bar{R}_v$ of dispersed phase were calculated by following equations:

$$\bar{R}_v = \frac{\sum n_i R_i^4}{\sum n_i R_i^3} \qquad (1)$$

The micro fibril morphology formed in the extrudates was studied by observing their cryofractured surface. For each sample the diameters of 100 fibrils were measured and the maximum, minimum and the average diameter were reported.

**RESULTS AND DISCUSSION**

*WAXD Results*

Figure 1 shows the WAXD patterns of PP/organoclay and COC/organoclay samples containing 5 wt. % organoclay (PP-N5 and COC-N5 in Table 1). The WAXD pattern of the organically modified montmorillonite (Cloisite 15A) is also inserted in this figure for comparison. As it can be seen the organoclay exhibits two characteristic peaks at 2θ=3.35º and 2θ=9.3º, corresponding to the interlayer spacing of the modified layers of montmorillonite and unmodified layers respectively. The results of interlayer spacing calculated from the angular location of peaks and Bragg's law are given in Table 2. Shifting of the peaks of organoclay to lower 2θ values can be considered as indication of melt interaction. Therefore, PP has greater ability in intercalating organoclay comparing to COC.

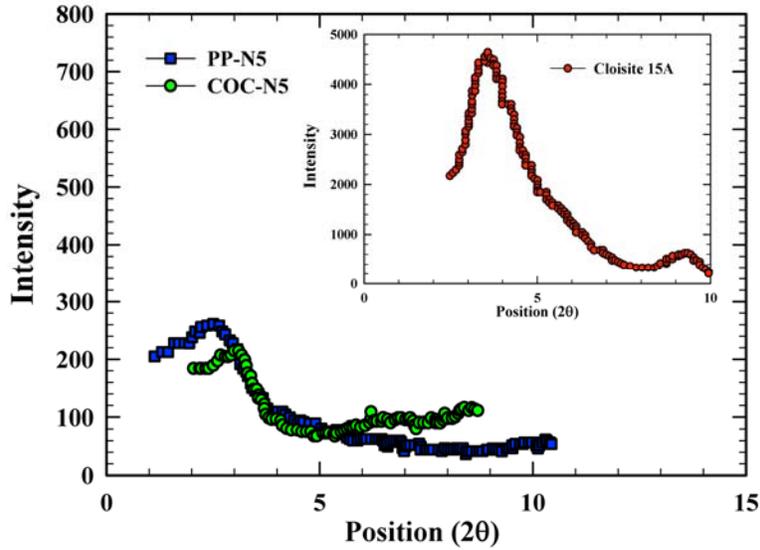

**Figure 1.** X-ray diffraction patterns of the organically modified montmorillonite (Cloisite 15A), PP-N5 and COC-N5.

**Table 2.** The inter-layer spacing values corresponding to the WAXD patterns shown in Figure1.

| sample | Position (2θ) | d-spacing (A) |
|---|---|---|
| Cloisite 15A | 3.35 | 26 |
| PP-N5 | 2.61 | 34 |
| COC-N5 | 3.01 | 29 |

*Linear Melt Viscoelastic Results*

Figure 2 presents the storage modulus, *G'* as a function of angular frequency for PP, COC, PP80/COC20 and COC80/PP20 samples. As it can be seen clearly, PP and COC exhibit terminal behavior at low frequencies. Interestingly, the low frequency storage modulus values of the blend samples are higher than their individual components with greater enhancement for PP80/COC20, as shown in the marked region. Slouf *et al.* have studied the phase disengagement mechanism in PP/COC blends and found that PP is not miscible with COC [38], thus the upraise in the low frequency storage modulus values for the blends PP80/COC20 and COC80/PP20 can be attributed to the strong interfacial adhesion between two phases. The elastic response of the droplet for PP80/COC20 sample, implying the capability of COC droplets for fibrillation could also be one of the reasons for the observed behavior.

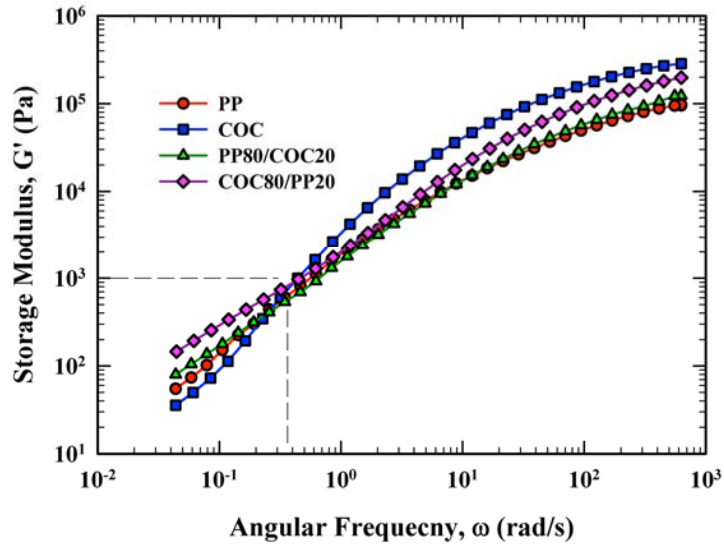

**Figure 2:** The storage modulus as a function of angular frequency for the samples PP, COC, PP80/COC20 and COC80/PP20.

Figure 3 demonstrates the variation of both storage modulus and complex viscosity with angular frequency of PP and PP-N5 samples. A pronounced low frequency plateau in storage modulus and a strong viscosity upturn were observed for PP-N5 in comparison with PP matrix, which is an indication of a three dimensional physical network formed between nanoparticles and/or nanoparticles and matrix chains. These results show great capacity of PP/PEgMA matrix in intercalation of nanoparticles. The results obtained from similar experiment for COC and COC-N5 are shown in Fig. 4. By comparing these results with those shown in Fig. 3, one may notice that PP has greater ability of intercalation of organoclay compared to COC in presence of PEgMA as compatibilizer.

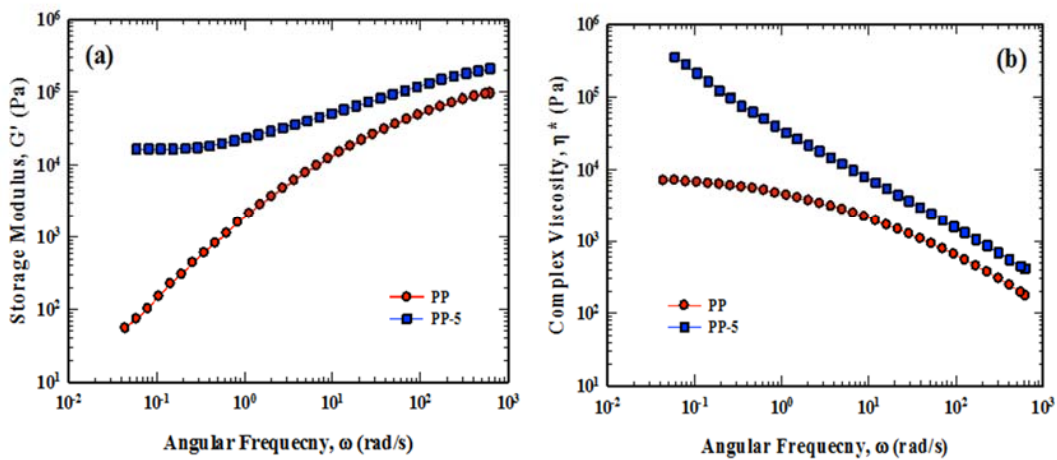

**Figure 3:** Variation of (a) storage modulus and (b) complex viscosity with angular frequency for PP and PP-N5.

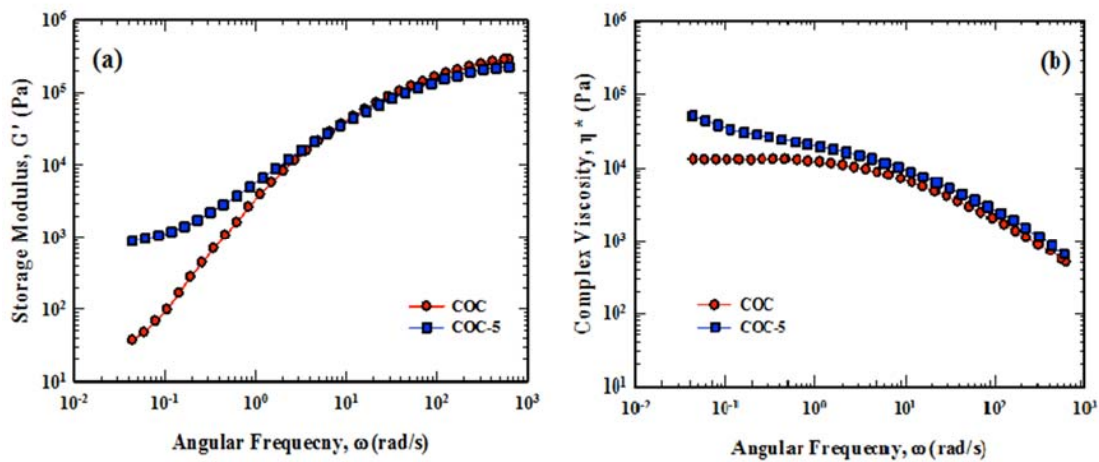

**Figure 4:** Variation of (a) storage modulus and (b) complex viscosity with angular frequency for COC and COC-N5.

In Figure 5, such results were shown for both PP-N5/COC and PP/COC-N5 blends, non-terminal behavior at low frequencies was observed for PP-N5/COC blend but not so for PP/COC-N5. The non-terminal behavior for PP-N5/COC indicates the formation of three-dimensional physical network in this blend. On the other hand, PP/COC-N5 shows similar behavior to PP matrix. These indications confirm the presence of major part of the organoclay tactoids and platelets in the PP matrix for PP-N5/COC and in the COC dispersed phase for PP/COC-N5. Discussed results confirm the importance of feeding procedure in determining the location of organocly.

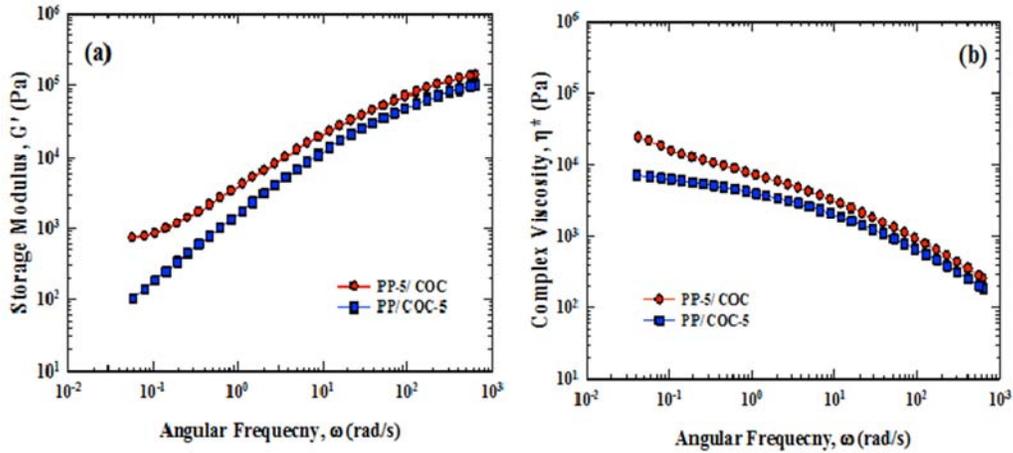

**Figure 5:** Variation of (a) storage modulus and (b) complex viscosity with angular frequency for PP-N5/COC and PP/COC-N5.

The effect of temperature on the storage modulus and complex viscosity of PP and COC is shown in Figure 6. It is very clear that PP is less thermo-sensitive than COC as the variation in storage modulus and complex viscosity of COC with temperature is very high compared to that of PP. This kind of behavior is very crucial in the cooling phase of the blend in the single extruder to save the generated morphology of PP/COC blends when COC is the dispersed phase.

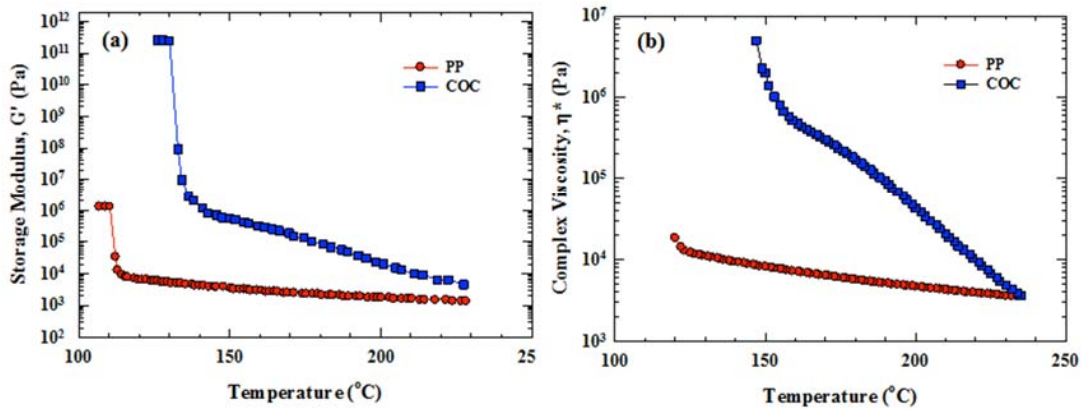

**Figure 6:** Effect of temperature on the storage modulus (a) and complex viscosity (b) of PP and COC.

*Morphology using SEM results*

*Morphology of Blends*

Figures 7a and 7b demonstrate the SEM micrographs of the cryofractured surface of the blends PP80/COC20, and COC80/PP20. As it can be seen, both blends exhibit a typical matrix-dispersed morphology where dispersed phase forms droplets finely dispersed in matrix due to the strong interfacial adhesion of the blend components. The average particle size measurements were performed on these SEM micrographs and the results are listed in Table 3.

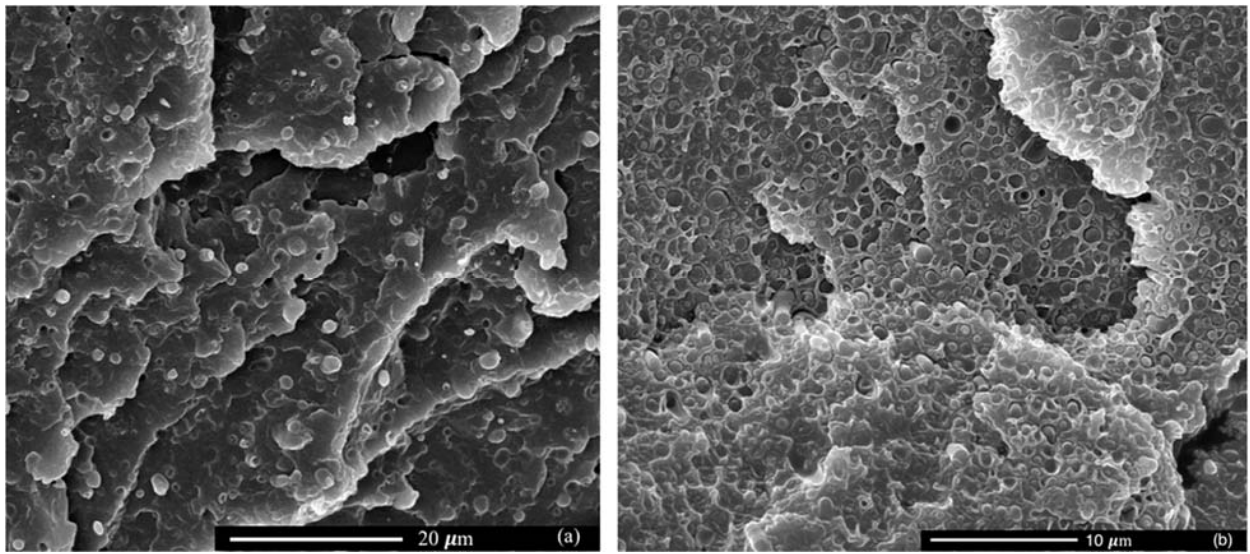

**Figure 7:** SEM micrographs of the cryofractured surface of the blend samples (a) PP80/COC20, (b) COC80/PP20.

**Table 3.** Volume average radius, $\bar{R}_v$ (μm), of PP80/COC20, PP-N5/COC and PP/COC-N5 blends.

| Sample | $\bar{R}_v$ (μm) |
|---|---|
| PP80/COC20 | 0.73 |
| COC80/PP20 | 0.4 |
| PP-N5/COC | 0.33 |
| PP/COC-N5 | 0.81 |

*Morphology of Unfilled Extrudates*

As can be noticed from Figure 7, morphology of the prepared blends PP80/COC20, and COC80/PP20 is established and stabilized via mixing in twin-screw extruder. As discussed in the experimental section, these blends were further extruded through a single-screw extruder to achieve microfibril formation of the dispersed phase. The obtained extrudates are referred as EPP80/COC20, and ECOC80/PP20, respectively, and their SEM micrographs are presented in Fig. 8. It is evident from Fig. 8. that the droplets formed in twin-screw extruder are converted to microfibrils by passing through the attached die of single-screw extruder. The corresponding fibril diameters of the dispersed components are listed in Table 4. Each sample is characterized by three values (the minimum, maximum, and average fibril diameter). These results suggest that in PP/COC blend system, unlike many other thermoplastic blends, which have only circular dispersed phase in the extruded samples, microfibril formation of droplets has occurred in the shear flow fields without any post processing such as fiber spinning [37, 38, 58]. The formation of fibrils in immiscible blends is mainly because of the enhanced viscous stresses (intending to deform the droplets) over interfacial stresses (trying to keep droplet shape unchanged) that causes increase in the droplet deformation and fibrillation. For ECOC80/PP20, high viscosity of matrix (COC) comparing to dispersed phase (PP) is in favor of viscous stresses and promotes fibrillation. Moreover, high glass transition temperature of COC matrix stabilizes PP fibrils and prevents PP from returning to circular form. For EPP80/COC20, although viscosity ratio is unfavorable for fibrillation, microfibrils of COC are observed. This can be attributed to the high elasticity of COC that prevents droplet breakup in the favor of fibrillation. Moreover, its high glass transition (140 ºC), leads to fast solidification of deformed droplets in the extrudates [37, 59-62]. The capillary stability and the capability of COC in fibril formation also have their contribution. This supports our argument on the rheology results, regarding the high elasticity of droplet deformation in PP80/COC20 blend at low frequencies.

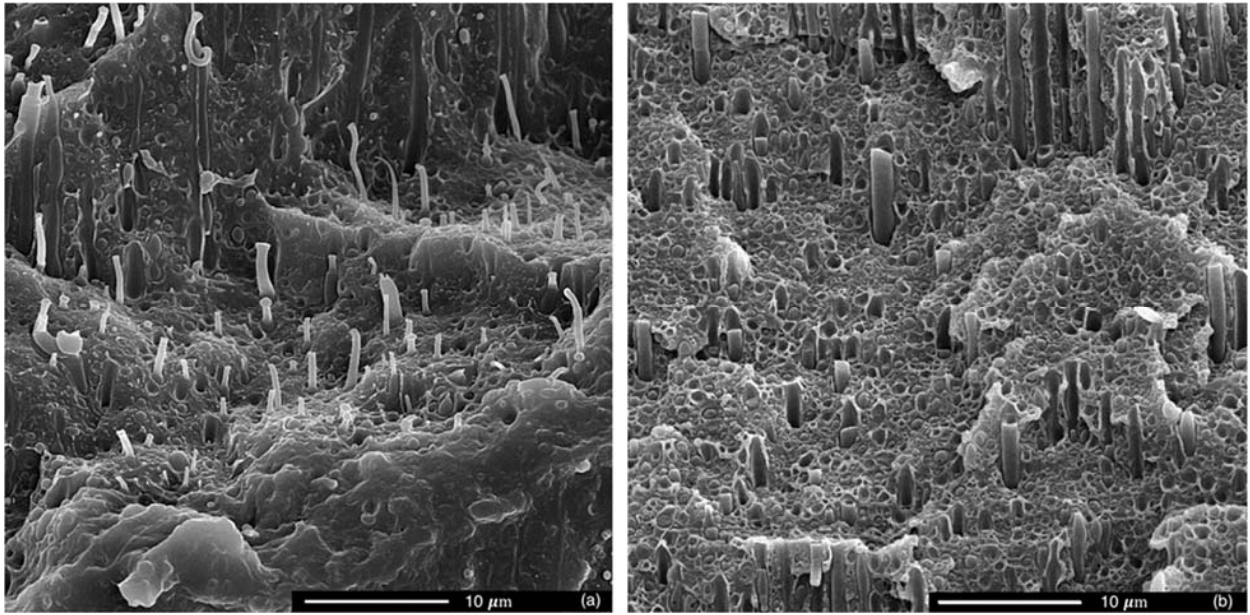

**Figure 8:** SEM micrographs of the extruded samples (a) EPP80/COC20, (b) ECOC80/PP20.

**Table 4:** Diameters of the fibers of the dispersed component in the matrix of the extruded polymer blends.

| Sample | Fibril diameter (µm) | | |
| --- | --- | --- | --- |
| | Minimum (µm) | Maximum (µm) | Average (µm) |
| EPP80/COC20 | 0.15 | 0.774 | 0.462 |
| ECOC80/PP20 | 0.42 | 1.329 | 0.874 |
| EPP-N5/COC | 0.11 | 0.526 | 0.318 |
| EPP/COC-N5 | 0.134 | 0.872 | 0.503 |

*Effect of Shear Rate*

The effect of shear rate on the droplet deformation and microfibril formation is further studied by employing three different screw speeds of 6, 60 and 120 rev/min corresponding to shear rates of 8, 72, and 143 $s^{-1}$, respectively in the single-screw extruder, for the blend PP80/COC20. The SEM micrographs of cryofractured surfaces of extrudates under these three different shear rates are shown in Fig. 9 and the corresponding fibril diameters are listed in Table 5. As can be noticed from the SEM micrographs and the values listed in Table 5, the droplet size decreases with increase in the shear rate. It is also evident that the microfibril formation of COC droplets increases with increasing shear rate; for shear rates 72 and 143 $s^{-1}$ the presence of COC fibrils in PP matrix is clearly observed.

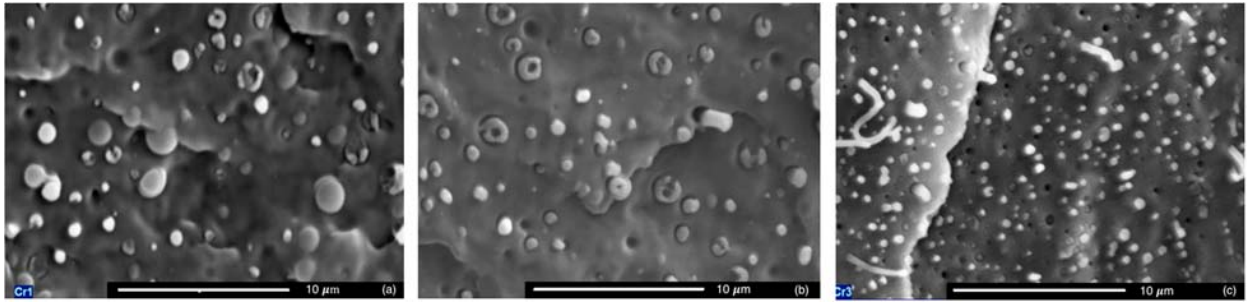

**Figure 9:** SEM micrographs of cryofractured surface of PP80/COC20 blend under these three different shear rates (a) 7.58 $s^{-1}$ (b) 72.10 $s^{-1}$ and (c) 142.71 $s^{-1}$.

**Table 5:** The particle size dimensions of extruded PP/COC blend under different shear rates.

| Sample | Shear rate (1/s) | Fibril diameter (μm) |
|---|---|---|
| EPP80/COC20 | 7.58 | 0.75 |
|  | 72.10 | 0.46 |
|  | 142.71 | 0.29 |

*Effect of Organoclay Partitioning*

As explained, in order to study the effect of organoclay partitioning, the nanocomposite samples with a fixed wt.% ratio of PP/COC (80/20) were prepared by two different mixing routes; in the first method, organoclay was located in PP matrix, and in the second one it was located in COC dispersed phase. The SEM micrographs of the cryofractured surfaces of these nanocomposites (PP-N5/COC, and PP/COC-N5) are illustrated in Figures 10a, and 10b. The volume average radius of dispersed phase ($\bar{R}_v$) for these samples are added to Table 3. $\bar{R}_v$ of PP-N5/COC, which was prepared by first melt mixing route, is much lower than that of unfilled blend (i.e. PP80/COC20). This is mainly due to the increased viscosity of PP matrix (due to the presence of organoclay), which leads to enhanced viscous forces that overcome the stabilizing interfacial tension. The increased viscous forces and reduced interfacial forces further increase the droplet deformation causing the break-up of COC droplets leading to reduced droplet size. Also, the presence of organoclay in the PP matrix and interface prevents the coalescence of COC droplets and thus reduces the droplet size [48, 49, 53, 54]. By close looking, one can recognize that some of the COC droplets are deformed into the form of fibrils, which is due to increased viscous forces from the matrix by presence of organoclay resulting in more droplet deformation. In case of PP/COC-N5 which was synthesized by second mixing route (organoclay was first melt mixed with COC), the droplet size of dispersed phase COC is higher than that of PP80/COC20, which can be explained by the reduction in droplet deformation due to the presence of organoclay in the dispersed phase COC and thus reducing the break-up of droplets. These observations are in agreement with the rheological measurements and confirm the importance of feeding method on morphology of prepared samples. These nanocomposite samples were further extruded to promote deformation of droplets. The produced extrudates are referred as EPP-N5/COC, and EPP/COC-N5 and Figure 11 presents their SEM micrographs. Formation of microfibrils is evident from this figure. The corresponding fibril diameters of the dispersed components are added to Table 4. Interestingly, presence of organoclay in PP matrix and COC dispersed phase reduces and increases the fibril diameters, respectively (Figs. 11a and 11b), which is consistent with our results for PP-N5/COC and PP/COC-N5 blends.

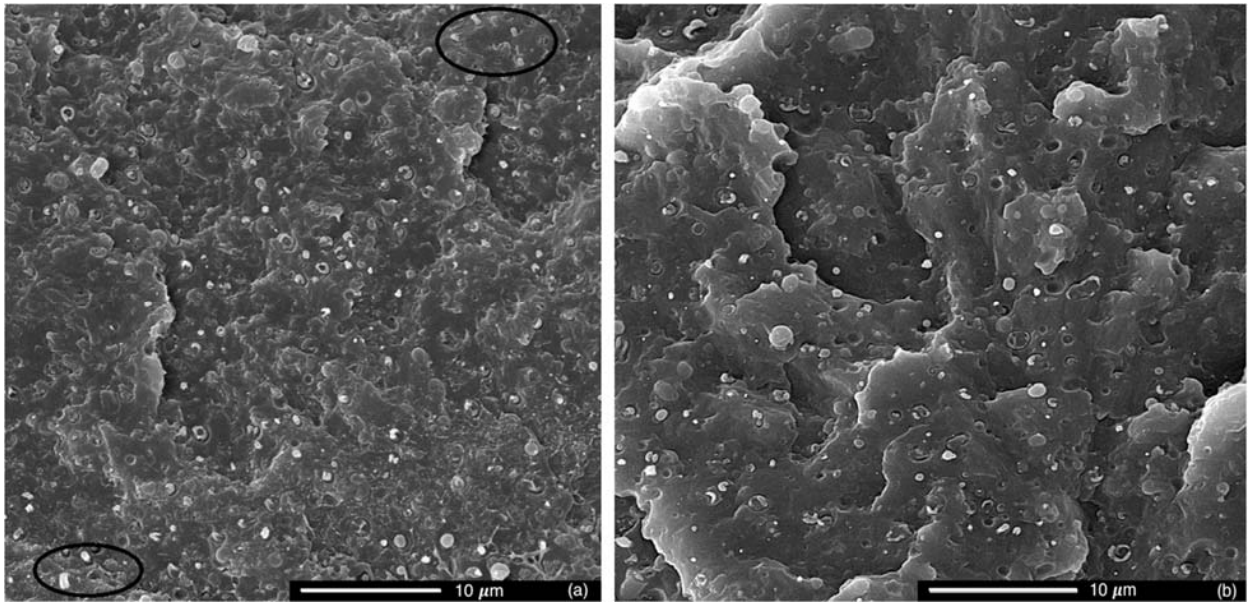

**Figure 10:** SEM micrographs of the cryofractured surface of the nanocomposites (a) PP-N5/COC, and (b) PP/COC-N5.

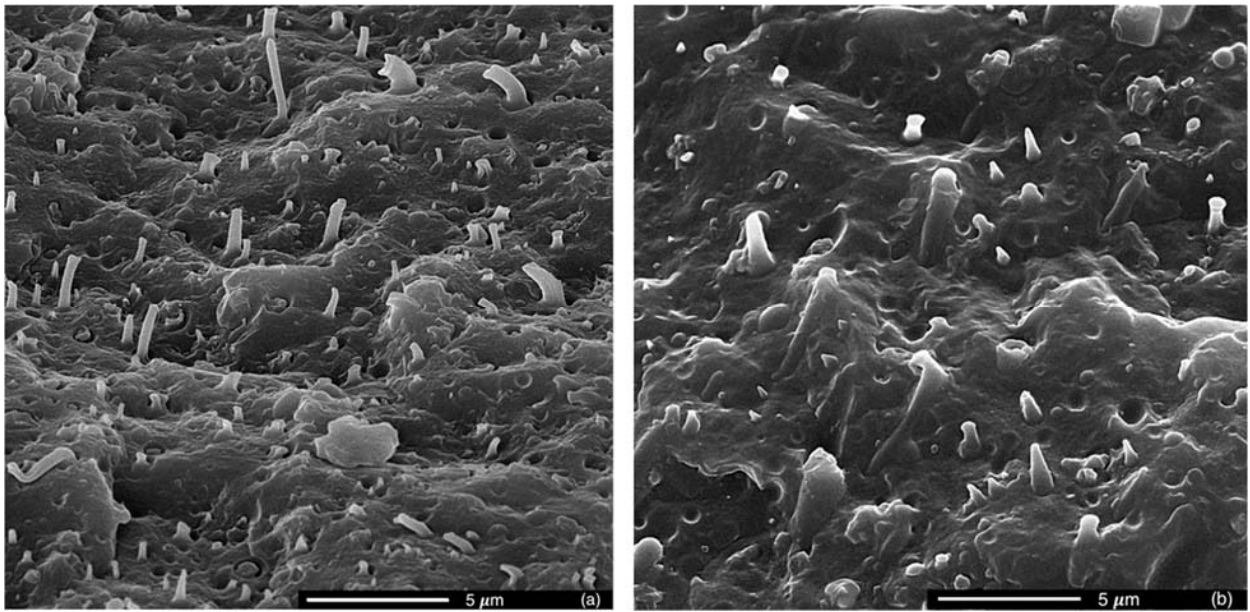

**Figure 11:** SEM micrographs of the extruded nanocomposites (a) EPP-N5/COC, and (b) EPP/COC-N5.

## MORPHOLOGY PREDICTION

The deformation and breakup of the dispersed phase depends on two dimensionless parameters [12], namely, viscosity ratio $p$, and capillary number, $Ca$. Capillary number is the ratio of the viscous stresses that favor the droplet deformation to the interfacial tension that keeps the droplet spherical (i.e., prevents break-up of droplets). The capillary number in simple shear flows is defined as:

$$Capillary\ number,\ Ca = \frac{\eta_m \dot{\gamma}}{\sigma_{12}/R} \qquad (2)$$

where $\eta_m$ is the matrix viscosity, $\dot{\gamma}$ the shear rate, $R$ radius of the droplet, and $\sigma_{12}$ the interfacial tension of polymer pairs. For capillary numbers above a critical value ($Ca_{crit}$), the droplet cannot sustain deformation any further, and breaks up eventually. The critical capillary number depends strongly on the type of flow i.e., shear or elongation and on the viscosity ratio ($p$). The critical capillary number ($Ca_{crit}$) in simple shearing flow is a function of viscosity ratio ($p$) and is given by [22]:

$$\log(Ca_{crit}) = -0.5060 - 0.0994\log p + 0.1240(\log p)^2 - 0.1150/(\log p - 0.6110) \qquad (3)$$

Utracki *et al.* [21, 22] defined a criteria for the dispersed droplets to either deform or breakup depending on the ratio of Ca to Ca$_{crit}$, i.e., Ca* in both shear and elongation flow fields. Many researchers have used this criteria to study/predict the morphology and fibril formation in polymer blends [62]. It has been well reported that the Ca* shall be more than 4, for the dispersed droplets to deform and form stable filaments or fibrils [62].

The interfacial tension between two components 1 and 2 can be obtained by using the well-known Owens and Wendt equation [63]:

$$\sigma_{12} = \sigma_1 + \sigma_2 - 2\sqrt{\sigma_1^d \sigma_2^d} - 2\sqrt{\sigma_1^p \sigma_2^p} \qquad (4)$$

Where the exponents $d$ and $p$ are the dispersive and polar contributions to the surface tension, respectively. The values of the surface tensions of PP and COC at 25°C are available in the literature [64, 65]. By using the temperature dependency relationship, the surface tension at 235 ºC (processing temperature) can be calculated. The extrapolated values of the surface tension at 235 ºC for components are given in Table 6. The calculated surface tension between PP and COC is 5.74 mN/m according to equation (4). For obtaining morphological and interfacial information on immiscible blends, the emulsion model of Palierne [66] has been used. The interfacial tension $\sigma_{12}$ for these blends is obtained by fitting their experimental and calculated complex modulus. The value of interfacial tension $\sigma_{12}$ is found to be 6.8 mN/m for PP-N5/COC and 7.1 mN/m for PP/COC-N5 samples.

**Table 6:** Values of surface tension of the components at 235ºC.

| Material | Total surface tension (mN/m) | Dispersive surface tension, $\sigma^d$ (mN/m) | Polar surface tension, $\sigma^p$ (mN/m) | $\frac{d\sigma}{dT}$ (mN/m.K) |
|---|---|---|---|---|
| PP | 17.5 | 17.5 | - | -0.06 |
| COC | 14.98 | 10.216 | 3.317 | -0.08 |
| Cloisite15A | 19.37 | 6.98 | 12.39 | - |

For PP-N5/COC, we assumed all the organoclay has remained in the matrix (i.e., no migration into the dispersed phase) and we used the viscosity of PP-N5 as the viscosity of the matrix. Similarly, for PP/COC-N5, all the organoclay content was assumed to be in the dispersed phase and the viscosity of COC-N5 was considered to be the viscosity of the dispersed phase. The Cox-Merz rule relates steady state non-linear material functions to linear viscoelastic properties as depicted by the following relationship.

$$\lim_{\dot{\gamma} \to 0} \eta(\dot{\gamma}) = \lim_{\omega \to 0} \eta'(\omega) \qquad (5)$$

We assumed that Cox-Merz rule is applicable to estimate the viscosities of the matrix and the dispersed phase and thus to calculate the viscosity ratio (*p*), at shear rate of 72 s$^{-1}$ which was used in our experiments. Given the viscosities of the matrix and the dispersed phase, capillary number and the critical capillary number values for the extruded samples EPP80/COC20,

ECOC80/PP20, EPP-N5/COC, and EPP/COC-N5 were calculated by using equations (2) and (3). These values are presented in Table 7.

**Table 7:** The capillary number and the critical capillary number values for the extruded polymer blends.

| Sample | $R_v$ (μm) | p | Ca | $Ca_{crit}$ | Ca* |
|---|---|---|---|---|---|
| EPP80/COC20 | 0.73 | 3.125 | 7.15 | 2.919 | 2.45 |
| ECOC80/PP20 | 0.4 | 0.32 | 12.25 | 0.476 | 25.75 |
| EPP-N5/COC | 0.33 | 1.25 | 6.82 | 0.512 | 13.32 |
| EPP/COC-N5 | 0.81 | 3.205 | 6.97 | 3.706 | 1.88 |

As shown in Figures 8 and 11, microfibrillation of the dispersed phase is observed for all the samples. The formation of microfibrils is expected for samples EPP-N5/COC and ECOC80/PP20, as the Ca* (i.e., the ratio of capillary number to the critical capillary number) values meet the criteria proposed by Utracki *et al.* [21, 22] (refer to Table 7). Even though the values of Ca* for samples EPP80/COC20 and EPP/COC-N5 are less than 4, microfibril formation of COC was observed. This can be attributed to the higher viscosity and elasticity of COC in comparison with PP, which leads to a more rigid structure in the molten state that inhibits the droplet breakup and favors the fiber formation. Also, sharp increase in the elasticity upon cooling and high glass transition of COC help saving the generated morphology when the COC is the dispersed phase (refer to Fig. 6). The parameters such as elasticity and glass transition of the dispersed phase are not considered in the Utracky's theory, and this may be the cause for the discrepancy observed between the morphology from SEM and prediction of Utracky's theory.

**CONCLUSIONS**

Rheological and morphological characteristics of PP/COC blends and PP/COC/organoclay have been studied using WXRD, linear viscoelastic measurements. The results indicated that the morphology of these blends, especially the size of the dispersed phase, depends strongly on organoclay partitioning, as evident from SEM micrographs as well. Temperature sweep measurements of storage modulus and complex viscosities of pure components PP and COC showed that COC is very thermosensitive in comparison with PP, which may be responsible for the generated morphology of PP/COC blends when the COC is in the dispersed phase during the cooling phase of the blend.

In addition, the effect of shear rate (in the single-screw extruder) on the microfibrillation capability of PP/COC blends has been investigated. The SEM micrographs of cryofractured surfaces of PP/COC blend at three different shear rates indicated that the droplet size decreases with increase in the shear rate. However, the polydispersity and microfibril formation of COC droplets increased with the shear rate. In contrast to many other polymers blends, it was observed that microfibril formation of COC droplets has occurred in the shear flow fields without any post processing. This is mainly because of the enhanced hydrodynamic stress over interfacial stress. The high glass transition of COC may also have considerable contribution.

The fibrillation of COC was found to be strongly dependent on the partitioning of organoclay. When the organoclay is in the PP matrix, the increased viscous forces over interfacial forces causes enhanced COC droplet deformation and thus leading to reduced droplet size. Presence of the organoclay in COC droplets, reduces the deformability and therefore fibrillation of COC.

Although the ratio of capillary number to the critical capillary number is less than 4, microfibril formation of COC is observed for few PP/COC extruded blends, in contrast to the fibril formation criteria defined by Utracki theory [21, 22]. This can be attributed to high elasticity and glass transition of the dispersed phase that are not included in the Utracky's theory.